\documentclass[article]{aa} 

\usepackage{natbib}
\usepackage{graphicx}
\usepackage{txfonts}

\begin{document}

\title{On the linear coupling between fast and slow MHD waves \\ due to line-tying
effects}

\author{J., Terradas$^{1}$, J., Andries$^{2,3}$, E., Verwichte$^{4}$}

\offprints{J. Terradas, \email{jaume.terradas@uib.es}}
\institute{$^1$Departament de F\'\i sica,
Universitat de les Illes Balears, E-07122, Spain, \email{jaume.terradas@uib.es}
\\$^2$Centre Plasma Astrophysics and Leuven Mathematical Modeling and
Computational Science Centre, Katholieke Universiteit Leuven, Leuven, B-3001,
Belgium, \email{jesse.andries@wis.kuleuven.be}
\\$^3$Centre for Stellar and Planetary Astrophysics, Monash University,
Victoria, 3800, Australia
\\$^4$Centre for Fusion, Space and Astrophysics, Department of Physics,
University of Warwick, Coventry CV4 7AL, UK, \email{Erwin.Verwichte@warwick.ac.uk}
}
{}

\date{Received / Accepted }

\abstract{Oscillations in coronal loops are usually interpreted in terms of
uncoupled magnetohydrodynamic (MHD) waves. Examples of these waves are  standing
transverse motions, interpreted as the kink MHD modes, and propagating slow
modes, commonly reported at the loop footpoints.}{Here we study a simple system
in which fast and slow MHD waves are coupled. The goal is to understand the
fingerprints of the coupling when boundary conditions are imposed in the
model.}{The reflection problem of a fast and slow MHD wave interacting with a
rigid boundary, representing the line-tying effect of the photosphere, is
analytically investigated. Both propagating and standing waves are analysed and
the time-dependent problem of the excitation of these waves is considered.}{An
obliquely incident fast MHD wave on the photosphere inevitably generates a slow
mode. The frequency of the generated slow mode at the photosphere is exactly the
same as the frequency of the incident fast MHD mode, but its wavelength is much
smaller, assuming that the sound speed is smaller than the Alfv\'en speed.}{ The
main signatures of the generated slow wave are density fluctuations at the loop
footpoints. We have derived a simple formula that relates the velocity amplitude
of the transverse standing mode with the density enhancements at the footpoints
due to the driven slow modes. Using these results it is shown that there are
possible evidences in the observations of the coupling between these two modes.}


\titlerunning{Coupling between fast and slow MHD waves}
\authorrunning{Terradas et al.}
\maketitle

\section{Introduction}

In coronal loops there is evidence of the presence of fast standing
magnetohydrodynamic (MHD) modes and slow (propagating and standing) MHD modes. 
Standing kink oscillations were first reported using TRACE by
\citet{aschetal99} and \citet{nakaetal99}. Later, similar observations were
analysed by, e.g., \citet{schrijbrown2000,aschetal02,schrijetal02}. There is
also clear evidence of the presence of propagating slow waves at loop footpoints
\citep[see][for a review]{demoortel09}. These slow waves are most likely due to
coupling between the underlying atmospheric layers since the dominant periods
tend to be around 5 min, suggesting a possible link with solar $p$-modes. Furthermore,
observations of standing slow modes in coronal loops, with periods largely above
5 min,  have been also reported by a number of authors, using different
instruments such as SOHO/SUMER
\citep[e.g.,][]{wangetal03b,wangetal03a,wangetal07}, Yohkoh/BCS
\citep[e.g.,][]{mariska05,mariska06} and more recently, Hinode/EIS
\citep[e.g.,][]{erdtaro08}. 

The theoretical interpretation of such a variety of oscillations is done, in most
of the cases, in terms of uncoupled MHD waves. On one hand, transverse
oscillations are identified as fast kink modes in the zero-$\beta$ approximation,
meaning that in the linear regime there is no longitudinal velocity along the
magnetic tube. On the other hand, the reported slow modes are associated to
acoustic modes, ignoring the coupling with the transverse motions. Although these
identifications are useful and simple, we have to bear in mind that under
realistic conditions MHD modes may couple. In general, the effect of gas
pressure, inhomogeneities or boundary conditions lead  to the coupling between
fast and slow modes. As we shall see, a clear example of coupling comes from the
line-tying condition, which amounts to considering the photosphere as providing a
complete reflection of any coronal disturbance impinging from above. This is
justified in most of the cases due to the large difference in densities between
the photosphere and corona but it obviously neglects the important role of the
transition region.

In this paper we study the simplest configuration in which fast and slow modes
are mixed to understand the basics of mode coupling. We use the simple idea
that a fast wave that is obliquely incident on a boundary will generate a slow
wave. As far as we know, this issue has been addressed in slightly different
contexts by, e.g., \citet{stein71,vasquez90,oliveretal92}. A remarkable work
about magnetohydrodynamic waves in coronal flux tubes including the line-tying
effect was carried out by \citet{goedhalb94}. In that work, it was clearly
stated that pure fast or pure slow modes do not exist in a line-tied coronal
loop. Here we follow a different approach to analyse this problem. Our interest
is in the effects of the application of line-tying conditions on fast and slow
modes and in the possible observational fingerprints of this coupling.






\section{Model and Dispersion relation}\label{model}

We consider first the simplest magnetic configuration where the magnetic field
points in the $z$-direction, the plasma density and pressure are constant, and
$v_{\rm A}>c_{\rm s}$ ($v_{\rm A}$ is the Alfv\'en speed and $c_{\rm s}$ the
sound speed). Since the
medium is homogeneous, we consider perturbations that are proportional to 
$e^{i\left(\omega t+k_x x + k_z
z\right)}$, meaning that waves propagate in the negative $x$ and $z$-directions.
In this configuration, the linearised MHD equations (see Appendix~\ref{app}) lead
to the well known dispersion relation for fast and slow MHD waves,
\begin{eqnarray}\label{omdisper} \omega^4-\left(k_x^2+k_z^2\right)\left(c_{\rm s}^2+v_{\rm
A}^2\right)\omega^2+k_z^2 \left(k_x^2+k_z^2\right)c_{\rm s}^2\,v_{\rm A}^2=0.
\end{eqnarray}
\noindent In our problem it is more convenient to assume that $\omega$ and $k_x$ are known
and solve for $k_z$. The reason is that the frequency of an incoming wave
remains constant in the reflection problem while the longitudinal wavenumber can
change. The dispersion relation  written as a biquadratic equation for $k_z$ is,
\begin{eqnarray}\label{kdisper}
c_{\rm s}^2\,v_{\rm A}^2\,k_z^4&-&\left[\left(c_{\rm s}^2+v_{\rm A}^2\right)\omega^2-c_{\rm s}^2\,v_{\rm A}^2\,k_x^2
\right]\,k_z^2\nonumber \\ &+& \omega^2\left[\omega^2-k_x^2\left(c_{\rm s}^2+v_{\rm A}^2\right)\right]=0.
\end{eqnarray}
We denote by $k_{\rm F}$ and $k_{\rm S}$ the fast and the slow longitudinal
wavenumbers that are solutions to the previous biquadratic equation, i.e.,
\begin{eqnarray}\label{kfdisper}
k_{\rm F}^2 &=&\frac{-B-\sqrt{B^2-4AC}}{2A},\\
k_{\rm S}^2 &=&\frac{-B+\sqrt{B^2-4AC}}{2A},\label{ksdisper}
\end{eqnarray}
where
\begin{eqnarray}\label{kdisper}
A&=&c_{\rm s}^2\,v_{\rm A}^2,\\
B&=&-\left[\left(c_{\rm s}^2+v_{\rm A}^2\right)\omega^2-c_{\rm s}^2\,v_{\rm A}^2\,k_x^2
\right],\\ 
C&=&\omega^2\left[\omega^2-k_x^2\left(c_{\rm s}^2+v_{\rm A}^2\right)\right].
\end{eqnarray}
These wavenumbers are associated to the same
frequency $\omega$ and same $k_x$, and can be approximated, using the small
plasma-$\beta$ assumption, 
by \citep[e.g.][]{oliveretal92}
\begin{eqnarray}\label{omfapp} k_{\rm F}^2&\simeq&\frac{\omega^2}{v_{\rm
A}^2}-k_x^2\left(\frac{c_{\rm s}^2}{v_{\rm A}^2}+1\right),\\
k_{\rm S}^2&\simeq&\frac{\omega^2}{c_{\rm s}^2}.\label{omsapp} \end{eqnarray}
These approximations will turn out to be useful in the following sections.
According to  Eq.~(\ref{omfapp}) the fast wavenumber may become purely imaginary for a
certain choice of the parameters, indicating that the wave is evanescent.
Hereafter,
we will restrict our analysis to propagating waves.

It can be seen from the linearised MHD equations that the velocity polarisation
for fast and slow modes is given by the following expression
\begin{eqnarray}\label{pol}
v_z=\frac{c_{\rm s}^2 k_x k_z}{\omega^2-k_z^2 c_{\rm s}^2} v_x,
\end{eqnarray}
where $k_z$ is either $k_{\rm F}$  or $k_{\rm S}$. Note that if we change the
direction of propagation along the field, $k_z$ changes to $-k_z$ and the
polarisation also changes sign. In our configuration and in the regime that we
are interested ($v_{\rm A}>c_{\rm s}$), since $k_{\rm F}<k_{\rm S}$, fast modes
are characterised by long wavelengths and by $v_x> v_z$, while slow modes have
short wavelengths and $v_x< v_z$. The slow modes have a larger compression since
the wavelength is small in comparison with the fast modes.

\section{Propagating waves: the reflection problem}\label{reflectan}

\subsection{Fast MHD wave reflection}

We consider the  most elementary problem to show the process of mode coupling due to
boundary conditions. A fast MHD wave travels downwards (in the negative $z$-direction)
and represents an incoming wave. This wave interacts with the photosphere (located at
$z=0$, where line-tying conditions are applied) and reflects (now travelling upwards).
The amplitude of the fast incoming wave is $F_{\rm I}$, while the amplitude of the
reflected fast wave is $F_{\rm R}$. To satisfy the boundary conditions a slow MHD wave,
also moving upwards, must be generated at $z=0$. The excited slow mode has an amplitude
$S_{\rm G}$. It is easy to write the velocity components using the polarisation of fast
and slow waves, given by Eq.~(\ref{pol}), and the proper sign of the longitudinal
wavenumber,

\begin{eqnarray} \label{v_xtot}
V_x=&F_{\rm I}&e^{i\left(\omega t+k_x x + k_{\rm F}
z\right)}+F_{\rm R}\,e^{i\left(\omega t+k_x x - k_{\rm F} z\right)}\nonumber \\ 
&-&S_{\rm G}\frac{\omega^2-k_{\rm S}^2 c_{\rm s}^2}{c_{\rm s}^2 k_x k_{\rm S}}\,e^{i\left(\omega t+k_x x - k_{\rm S}
z\right)},\\
V_z=&F_{\rm I}&\frac{c_{\rm s}^2 k_x k_{\rm F}}{\omega^2-k_{\rm F}^2
c_{\rm s}^2}\,e^{i\left(\omega t+k_x x + k_{\rm F} z\right)}-F_{\rm R}\frac{c_{\rm s}^2 k_x
k_{\rm F}}{\omega^2-k_{\rm F}^2 c_{\rm s}^2}\,e^{i\left(\omega t+k_x x - k_{\rm F} z\right)}\nonumber
\\  &+&S_{\rm G}\,e^{i\left(\omega t+k_x x - k_{\rm S} z\right)}.\label{v_ztot}  \end{eqnarray}

\noindent Now boundary conditions are imposed at $z=0$, representing the location
of the photosphere. We use line-tying conditions, i.e., $V_x(z=0)=V_z(z=0)=0$.
According to Eqs.~(\ref{v_xtot}) and (\ref{v_ztot}) the following conditions
must be satisfied, \begin{eqnarray} F_{\rm I}+F_{\rm R}-S_{\rm G}\frac{\omega^2-k_{\rm S}^2 c_{\rm s}^2}{c_{\rm s}^2
k_x k_{\rm S}}&=&0,\\ \nonumber F_{\rm I}-F_{\rm R}+S_{\rm G}\frac{\omega^2-k_{\rm F}^2 c_{\rm s}^2}{c_{\rm s}^2 k_x
k_{\rm F}}&=&0. \end{eqnarray} The solution in terms of the amplitude of the incoming
wave (which is an arbitrary parameter) is   \begin{eqnarray}
\frac{F_{\rm R}}{F_{\rm I}}&=&
\frac{\left(\omega^2-k_{\rm S}^2
c_{\rm s}^2\right)k_{\rm F}+
\left(\omega^2-k_{\rm F}^2 c_{\rm s}^2\right)k_{\rm S}}{\left(\omega^2-k_{\rm S}^2 c_{\rm s}^2\right)k_{\rm F}-
\left(\omega^2-k_{\rm F}^2 c_{\rm s}^2\right)k_{\rm S}},\label{frt}\\
\frac{S_{\rm G}}{F_{\rm I}}&=&
\frac{2 c_{\rm s}^2 k_x k_{\rm S} k_{\rm F}}{\left(\omega^2-k_{\rm S}^2 c_{\rm s}^2\right)k_{\rm F}-
\left(\omega^2-k_{\rm F}^2 c_{\rm s}^2\right)k_{\rm S}}.\label{srt}\end{eqnarray} 

\noindent The coefficients depend on $\omega$, $k_x$ (same for the three waves)
and on longitudinal wavenumbers, $k_{\rm F}$ and $k_{\rm S}$, which can also be
expressed in terms of $\omega$, $k_x$ by using Eqs.~(\ref{kfdisper})
and~(\ref{ksdisper}).

\begin{figure}[!ht]
\center{\includegraphics[width=9cm]{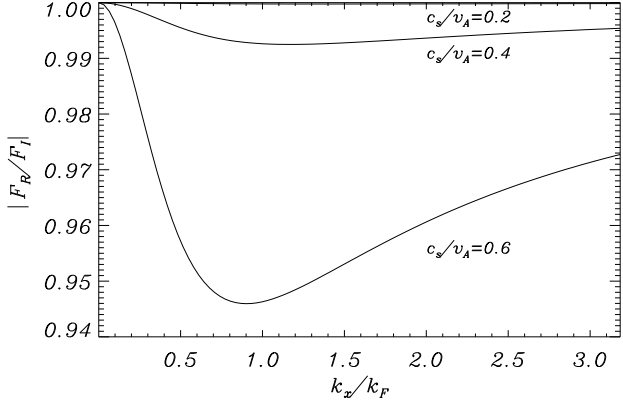}} \caption{\small
Amplitude of reflected fast mode given by Eq.~(\ref{frt}) as a function of the horizontal wavenumber for three
different values of the sound speed. 
}
\label{frplot} \end{figure}

When there is no coupling between fast and slow modes (i.e., when $k_x=0$)  we
have that $\left| F_{\rm R}/F_{\rm I} \right|=1$ because $\omega^2=k_{\rm S}^2
c_{\rm s}^2$, while $S_{\rm G}/F_{\rm I}=0$. Hence, there is a complete
reflection of the incoming fast mode while the slow mode is absent. When there is
coupling, the reflection coefficient for the slow mode is always different from
zero, meaning that the incoming fast MHD wave inevitably generates a slow mode at
the boundary. In Figure~\ref{frplot} the reflection coefficient of the fast wave,
i.e., $F_{\rm R}/F_{\rm I}$, is plotted as a function of $k_x$ for different
values of the sound speed. The coefficient is close to unity for  values of sound
speed low in comparison with the Alfv\'en speed, meaning that reflection is
almost total. However, for relatively large values of $c_{\rm s}/v_{\rm A}$ the
curves clearly show a minimum indicating that the reflection of the fast wave is
less efficient.

\begin{figure}[!ht] \center{\includegraphics[width=9cm]{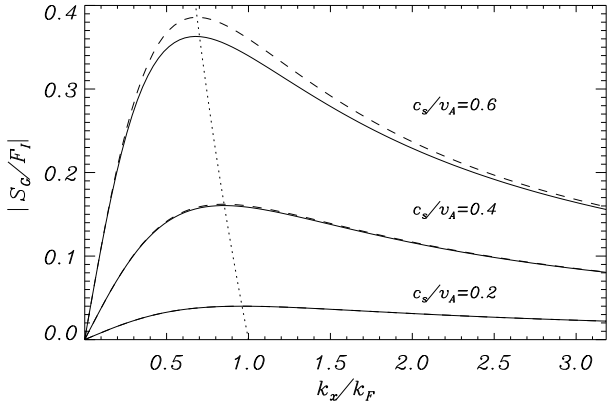}} \caption{\small
Amplitude of the generated slow mode given by Eq.~(\ref{srt}) as a function of
the horizontal wavenumber for three different values of the sound speed. The
dashed line represents the approximation given by Eq.~(\ref{srapprox1}) while
the dotted line shows the position of the maxima according to Eqs.~(\ref{kxmax})
and~(\ref{srapproxmax}). 
} \label{srplot}
\end{figure}

In Figure~\ref{srplot} the reflection coefficient of the slow wave, i.e., $S_{\rm
G}/F_{\rm I}$,  is plotted. The behaviour of the reflected slow mode amplitude is
basically opposite of the reflected fast wave. The curves show a maximum at the
locations where the generation of the slow mode is more efficient. Again the
value at the maximum strongly depends on the ratio of the sound speed to the
Alfv\'en speed, and corresponds to a $k_x$ which is very similar to $k_{\rm F}$.
The location of the maximum can be analytically determined from Eq.~(\ref{srt})
and using the approximation  of the slow mode frequency, $\omega^2 \simeq k_{\rm
S}^2\,c_{\rm s}^2$,
\begin{eqnarray}\label{srapprox}
\frac{S_{\rm G}}{F_{\rm I}}\simeq\frac{2 k_x
k_{\rm F}}{k_{\rm F}^2-k_{\rm S}^2}. \end{eqnarray} 
Since the frequency is fixed, we have from
the approximate expressions ~(\ref{omfapp})-(\ref{omsapp}) for $k_{\rm F}^2$ and $k_{\rm
S}^2$ the following relation
\begin{eqnarray}\label{ks-kf} k_{\rm S}^2 \simeq
k_x^2\left(1+\frac{v_{\rm A}^2}{c_{\rm s}^2}\right)+k_{\rm F}^2\frac{v_{\rm A}^2}{c_{\rm s}^2}.
\end{eqnarray}
Inserting this expression in Eq.~(\ref{srapprox}) we find that
\begin{eqnarray}\label{srapprox1}
\frac{S_{\rm G}}{F_{\rm I}}\simeq\frac{2 c_{\rm s}^2 k_x
k_{\rm F}}{k_{\rm F}^2\left(c_{\rm s}^2-v_{\rm A}^2\right)-k_x^2\left(c_{\rm s}^2+v_{\rm A}^2\right)}.\end{eqnarray}
This is a good approximation to the slow mode reflection coefficient (see dashed
line in Figure~\ref{srplot}), at least for the case $c_{\rm s} \ll v_{\rm A}$.
It is used to determine the location of the maxima of the curves in
Figure~\ref{srplot}. Imposing that the derivative of Eq.~(\ref{srapprox1}) is
equal to zero it is found that the  maximum corresponds to 
\begin{eqnarray}\label{kxmax} {k_x}_{\rm max} \simeq
k_{\rm F} \sqrt{\frac{v_{\rm A}^2-c_{\rm s}^2}{v_{\rm A}^2+c_{\rm s}^2}},
\end{eqnarray}
and the value of the reflection coefficient for this horizontal wavenumber  is
\begin{eqnarray}\label{srapproxmax}
\left|\frac{S_{\rm G}}{F_{\rm I}}\right|_{\rm max}\simeq\frac{c_{\rm s}^2}{\sqrt{v_{\rm A}^4-c_{\rm s}^4}}.
\end{eqnarray} 

\noindent In Figure~\ref{srplot} the location and value of the maximum is
represented by a dotted line. There is excellent agreement between the
approximation and the location of the maximum calculated using the original
expression for the reflection coefficient of the slow mode (Eq.~(\ref{srt})).
From Eq.~(\ref{srapproxmax}) we see that the maximum amplitude of the generated
slow wave depends only on the $\beta$ of the plasma.

Once we know the coefficients it is straight forward to calculate magnitudes of
interest such as the density perturbation of the slow reflected wave as a
function of amplitude of the incoming fast wave. As we will discuss later, these
magnitudes can be directly related to real observations. From the continuity
equation (see Appendix~\ref{app}) the density perturbation for the
slow mode is
\begin{eqnarray}\label{denswave} \left|\frac{\rho_1}{\rho}\right|=\frac{1}{\omega}\left(k_x
V_x+k_z V_z\right)=\frac{k_{\rm S}}{\omega}\left(1-\frac{\omega^2-k_{\rm S}^2 c_{\rm s}^2} 
{k_{\rm S}^2 c_{\rm s}^2}\right){S_{\rm G}}. \end{eqnarray}
Using again the fact that $\omega^2 \simeq k_{\rm S}^2\,c_{\rm s}^2$ and Eq.~(\ref{srapproxmax})
we find
\begin{eqnarray}\label{denstslowmax}
\left|\frac{\rho_1}{\rho}\right|_{\rm max}\simeq\frac{c_{\rm s} F_{\rm I}}{\sqrt{v_{\rm A}^4-c_{\rm s}^4}}.
\end{eqnarray}
This equation relates the maximum density perturbation associated to the
generated slow mode with the velocity amplitude of the incident fast wave.


\subsection{The time-dependent problem}

The time-dependent problem is numerically solved to demonstrate the mode coupling
described in the previous section. The temporal evolution of the waves and
specially their interaction with the boundaries provides a clear picture of the
coupling and complements the analytical results.

In this numerical experiment a pulse in the $v_x$ component is generated at
$t=0$. This pulse has the following form 
\begin{eqnarray} \label{v_xpropag}
v_x(z,t=0)&=&v_0 \cos\left(k z\right)\,e^{-\left(\frac{z-z_0}{a}\right)^2},\\
v_z(z,t=0)&=&0,\label{v_zpropag} \end{eqnarray}
and the rest of the perturbed
variables are set to zero. This particular profile has the property that
represents a localised wave packet but with a rather well defined wavelength
(basically given by $\lambda=2\pi/k$). To show the interaction of the fast mode
with a single boundary, rigid conditions are applied at $z=0$
($v_x(z=0)=v_z(z=0)=0$) while open conditions are imposed at $z=L$. The
linearised time-dependent MHD equations (see Appendix~\ref{app}) are numerically
solved using standard finite differences techniques.

\begin{figure}[!h]
\center{\includegraphics[width=7cm]{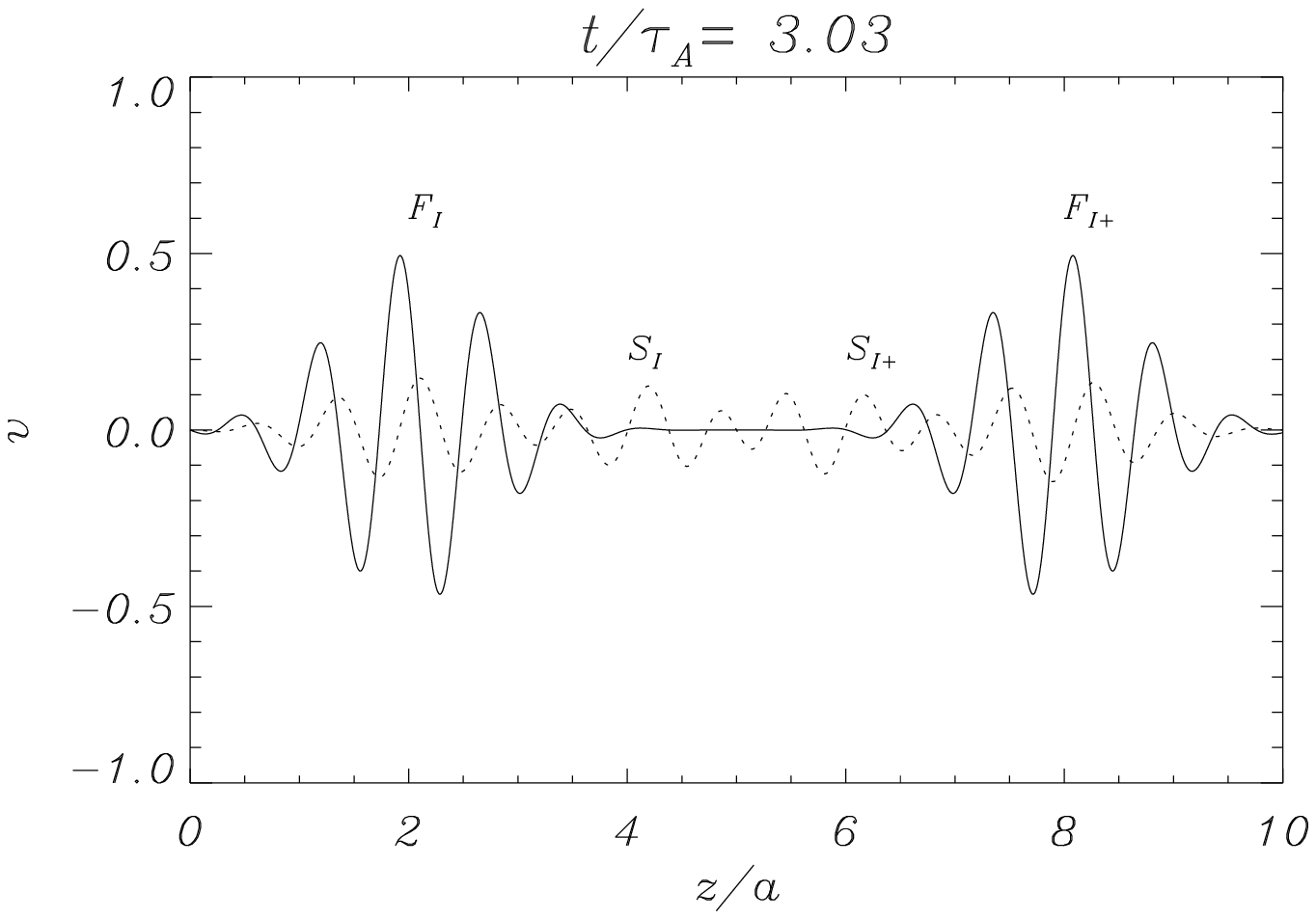}}
\center{\includegraphics[width=7cm]{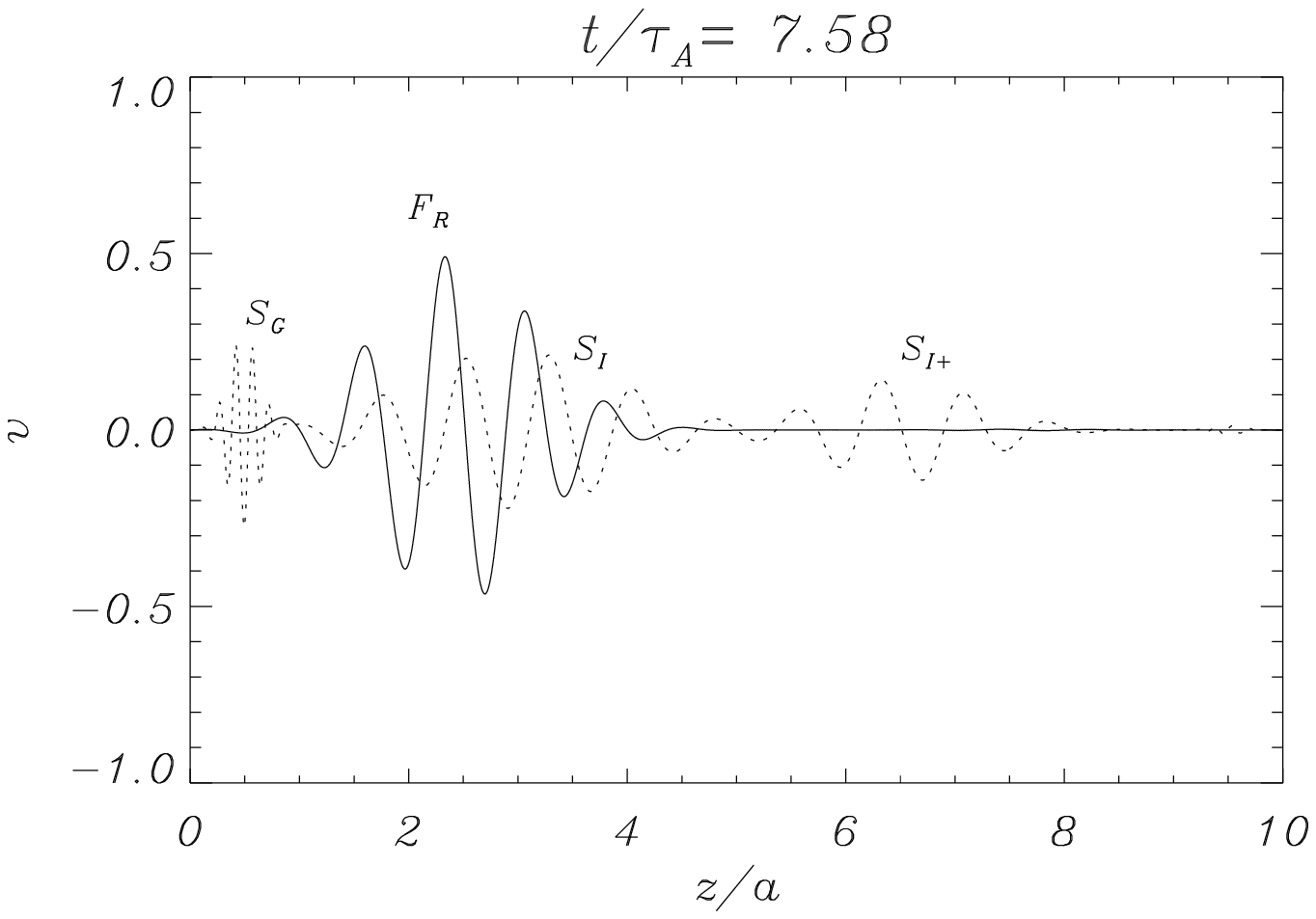}}
\center{\includegraphics[width=7cm]{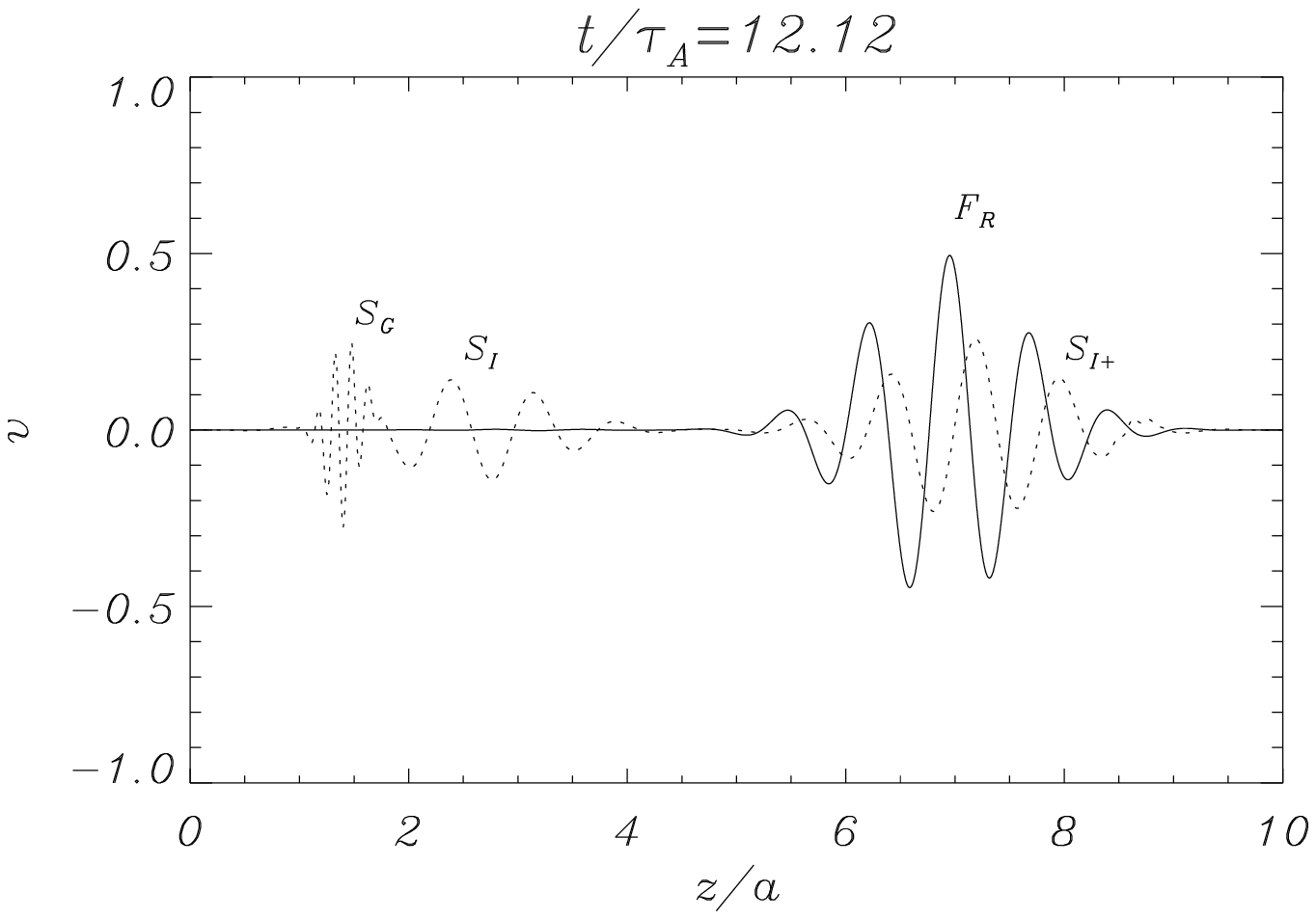}}
\caption{\small Time evolution of the propagating disturbance given by
Eqs.~(\ref{v_xpropag})-(\ref{v_zpropag}). The continuous line represents the $v_x$ component while
the $v_z$ component is plotted with a dotted line (its amplitude has been
multiplied by a factor 40 for visualisation purposes). Fast and slow modes travelling to the left are denoted as $F_{\rm I}$ and
$S_{\rm I}$, $F_{I+}$ and $S_{I+}$ move to the right, and $S_{\rm G}$ is the generated wave
at $z=0$ (moving to the right). For this simulation $z_0/a=5$, $k a=2\pi/0.75$,
$v_0/v_{\rm A}=1$, $k_x a=1.5$ and $c_{\rm s}/v_{\rm A}=0.2$. In this plot the
time has been normalised to the
characteristic time scale, $\tau_{\rm A}=v_{\rm A}/a$.} 
\label{reflectplot} \end{figure}

The temporal evolution of the two velocity components is plotted in 
Figure~\ref{reflectplot}. The initial disturbance excites the fast mode but also
the slow mode since the initial perturbation does not satisfy the velocity
polarisation relation, given by Eq.~(\ref{pol}). The excited fast and slow
modes split in two identical modes propagating in opposite directions (top
panel). The fast and slow waves travelling to the left are denoted as $F_{\rm
I}$ and $S_{\rm I}$, respectively, while $F_{\rm I+}$ and $S_{\rm I+}$ move to
the right. Once the $F_{\rm I}$ mode reaches the rigid boundary ($z=0$), it gets
reflected  (see middle panel) with an amplitude $F_{\rm R}$ and a slow mode is
generated (see dotted line), i.e., the $S_{\rm G}$ mode according to the
notation introduced above. The slow wave packet has a typical wavelength much
smaller than the fast wavelength ($k_{\rm F}\ll k_{\rm S}$). At the other
boundary ($z=L$) the fast mode $F_{\rm I+}$ leaves the system due to the
transparent boundary conditions. At later times (bottom panel) the $S_{\rm G}$
and $S_{\rm I}$ modes, moving in opposite directions, start to superpose, while
the $F_{\rm R}$ and $S_{\rm I+}$, travelling in the same direction also
interfere.  

The amplitudes of reflected fast and slow waves estimated from the
time-dependent problem agree quite well with the calculations based on the
analytical reflection coefficients given by Eqs.~(\ref{frt})-(\ref{srt}). Note
that these equations were derived for purely sinusoidal and monochromatic waves
but the initial packet is well represented by a dominant frequency (wavelength),
and this is the reason of the agreement with the analytical results.

\subsection{Slow MHD wave reflection}

Although we are mainly interested in the fast reflection problem, for completeness the slow mode reflection problem is also studied. The
difference with respect to the fast MHD reflection problem is that reflection of the
slow MHD mode at the boundary generates a fast MHD mode. The velocity components for
this problem are 
\begin{eqnarray}
\label{v_xtotslow} V_x=&\bar{S}_{\rm I}&\frac{\omega^2-k_{\rm S}^2 c_{\rm
s}^2}{c_{\rm s}^2 k_x k_{\rm S}}\,e^{i\left(\omega t+k_x x +k_{\rm S}
z\right)}-\bar{S}_{\rm R}\frac{\omega^2-k_{\rm S}^2 c_{\rm s}^2}{c_{\rm s}^2 k_x
k_{\rm S}}\,e^{i\left(\omega t+k_x x - k_{\rm S} z\right)}\nonumber \\ &+&
\bar{F}_{\rm G}\,e^{i\left(\omega t+k_x x - k_{\rm F} z\right)},\nonumber \\ \\
V_z=&\bar{S}_{\rm I}&\,e^{i\left(\omega t+k_x x - k_{\rm S}
z\right)}+\bar{S}_{\rm R}\,e^{i\left(\omega t+k_x x - k_{\rm S}
z\right)}\nonumber \\   &-&\bar{F}_{\rm G}\frac{c_{\rm s}^2 k_x k_{\rm
F}}{\omega^2-k_{\rm F}^2 c_{\rm s}^2}\,e^{i\left(\omega t+k_x x - k_{\rm F}
z\right)}.\label{v_ztotslow}  \end{eqnarray}

\noindent Applying rigid boundary conditions the following coefficients for the
reflected slow mode and the generated fast mode  are found
\begin{eqnarray}
\frac{\bar{S}_{\rm R}}{\bar{S}_{\rm I}}&=&
\frac{\left(\omega^2-k_{\rm S}^2
c_{\rm s}^2\right)k_{\rm F}+
\left(\omega^2-k_{\rm F}^2 c_{\rm s}^2\right)k_{\rm S}}{\left(\omega^2-k_{\rm S}^2 c_{\rm s}^2\right)k_{\rm F}-
\left(\omega^2-k_{\rm F}^2 c_{\rm s}^2\right)k_{\rm S}},\label{frtslow}\\
\frac{\bar{F}_{\rm G}}{\bar{S}_{\rm I}}&=&
\frac{2}{c_{\rm s}^2 k_x}\frac{\left(\omega^2-k_{\rm S}^2 c_{\rm s}^2\right)\left(\omega^2-k_{\rm F}^2 c_{\rm s}^2\right)k_{\rm S}}{\left(\omega^2-k_{\rm S}^2 c_{\rm s}^2\right)k_{\rm F}-
\left(\omega^2-k_{\rm F}^2 c_{\rm s}^2\right) k_{\rm S}}.\label{srtslow}\end{eqnarray}

\noindent It is worth to note that the reflection coefficient of the slow mode
is exactly the same as the reflection coefficient of the fast mode given by
Eq.~(\ref{frt}).

\section{Standing waves}

Once we understand the reflection of a propagating fast and a propagating slow wave at a
rigid boundary we extend our analysis to a more realistic physical situation, i.e., the
standing fast wave problem that is commonly reported by TRACE. 

There are several ways to study this problem, and the perturbation method is
one of them. Under this approach, we know that for zero-$\beta$ we have a pure
standing fast wave, with only a $v_x$ velocity component, that satisfies the
boundary conditions. When $\beta$ is different from zero but small, a $v_z$
component is introduced, due to Eq.~(\ref{pol}), and even worse, this component
does not satisfy the boundary condition. In order to compensate,  we have to
add to the system a slow wave perturbation (with a dominant $v_z$ in comparison
with $v_x$), which combined with the $v_z$ of the fast, will satisfy the
boundary conditions jointly. Now the $v_x$ introduced by the slow mode is
irrelevant since it is higher order in $\beta$ (which is assumed to be small). 

Another way to analyse this problem is to solve the full standing problem. This
was already done by \citet{oliveretal92} and \citet{goedhalb94} and it is based
on the superposition of the fast reflection problem plus the slow reflection
problem. The perturbation scheme and the full eigenmode problem give similar
results, except where the solutions are completely mixed, i.e., when the
slow mode component is not small in comparison with the fast component. This
takes place around the ``avoided crossings" in the dispersion diagram
\citep[see][for further details]{oliveretal92}.

The advantage of the perturbation scheme is that we can still obtain simple
approximations for the velocity and density changes around the footpoints. Let
us assume that an initial disturbance mostly excites the standing pattern in
the $v_x$ component. This standing wave is composed by the superposition of two
identical propagating waves travelling in opposite directions. We concentrate on
the fundamental mode, having a maximum at $z=L/2$ and a node of the velocity at
the boundaries. A
propagating slow wave is induced by the reflection at the boundaries of the
incident propagating fast wave that forms part of the standing fast pattern.
Hence, if the transverse velocity amplitude of the standing fast wave at the
loop apex is $V_0$ the amplitude of the incoming fast wave is $V_0/2$ (the
reflected fast wave will also have an amplitude of $V_0/2$ in our
approximation). Once we know this amplitude it is easy to estimate the velocity
or density fluctuations associated to the slow mode using the results of the
propagation problem. According to the analysis
performed in the previous section (see Eqs.~(\ref{srapproxmax}]) 
and (\ref{denstslowmax})) 
\begin{eqnarray}\label{srapproxmax1}
\left|v_z\right|_{\rm max}\simeq\frac{c_{\rm s}^2}{\sqrt{v_{\rm A}^4-c_{\rm
s}^4}} \frac{V_0}{2}, \end{eqnarray}  and \begin{eqnarray}\label{denstslowstand}
\left|\frac{\rho_1}{\rho}\right|_{\rm max}\simeq\frac{c_{\rm s}}{\sqrt{v_{\rm
A}^4-c_{\rm s}^4}}\frac{V_0}{2}. \end{eqnarray} 
These very simple expressions
give the order of the maximum velocity and density perturbations associated to the slow
reflected mode as a function of the transverse velocity amplitude at the loop
apex and the characteristic speeds of the configuration. When $c_{\rm s} \ll
v_{\rm A}$ the velocity perturbation reduces to 
\begin{eqnarray}\label{srapproxmax1app} \left|v_z\right|_{\rm
max}\simeq\frac{c_{\rm s}^2}{v_{\rm A}^2} \frac{V_0}{2}=\frac{1}{4}\gamma \beta V_0, \end{eqnarray}
while 
the density perturbation is  \begin{eqnarray}\label{denstslowstandapp}
\left|\frac{\rho_1}{\rho}\right|_{\rm max}\simeq\frac{c_{\rm s}}{v_{\rm
A}^2}\frac{V_0}{2} =\frac{1}{4}\gamma \beta \frac{V_0}{c_{\rm s}}. \end{eqnarray}

\noindent It is worth to calculate the order of magnitude of velocity and
density fluctuations around the footpoint. Let us take typical values: $v_{\rm
A}=800 \,\rm km\,s^{-1}$, $c_{\rm s}=200\,\rm km\,s^{-1}$, $V_0=80\,\rm
km\,s^{-1}$. Thus, the velocity perturbation according to
Eq.~(\ref{srapproxmax1}) is $\left|v_z\right|_{\rm max}\simeq 2.5\, \rm km\,
s^{-1}$. Using Eq.~(\ref{denstslowstand}) we find that the density fluctuation
associated to the slow mode is $\left| \rho_1 / \rho \right|_{\rm max}\simeq
1.25\%$. These density perturbations, although small, should be detectable with
the current instruments. Obviously, larger sound speeds and or small Alfv\'en
velocities would result in an increase of velocity and density fluctuations.

The previous estimations are based on the assumption
that the excited slow modes do not have time to reach the opposite footpoint and
reflect. If this is the case, the problem is more difficult since the slow modes
interfere and the velocity and density changes can be larger than those given by
Eqs.~(\ref{srapproxmax1})-(\ref{denstslowstand}). It is also important to remark
that the density perturbation associated to the fast standing mode has the same
profile as the velocity (the $v_x$ component), therefore the fast density
perturbation is zero at the footpoints (and maximum at the apex), this explains
why it is enough to consider only the density changes associated to the excited
slow modes around the footpoints.

\subsection{The time-dependent problem}

Now the time-dependent problem is solved. The initial perturbation, representing
the excitation of a mainly fast standing wave,  has the following profile, 
\begin{eqnarray} \label{v_xstanding} v_x(z,t=0)&=&v_0\sin\left(k z\right),\\
v_z(z,t=0)&=&0.\label{v_zstanding} \end{eqnarray}

\subsubsection{Rigid boundary conditions}

\begin{figure}[!h]
\center{\includegraphics[width=7cm]{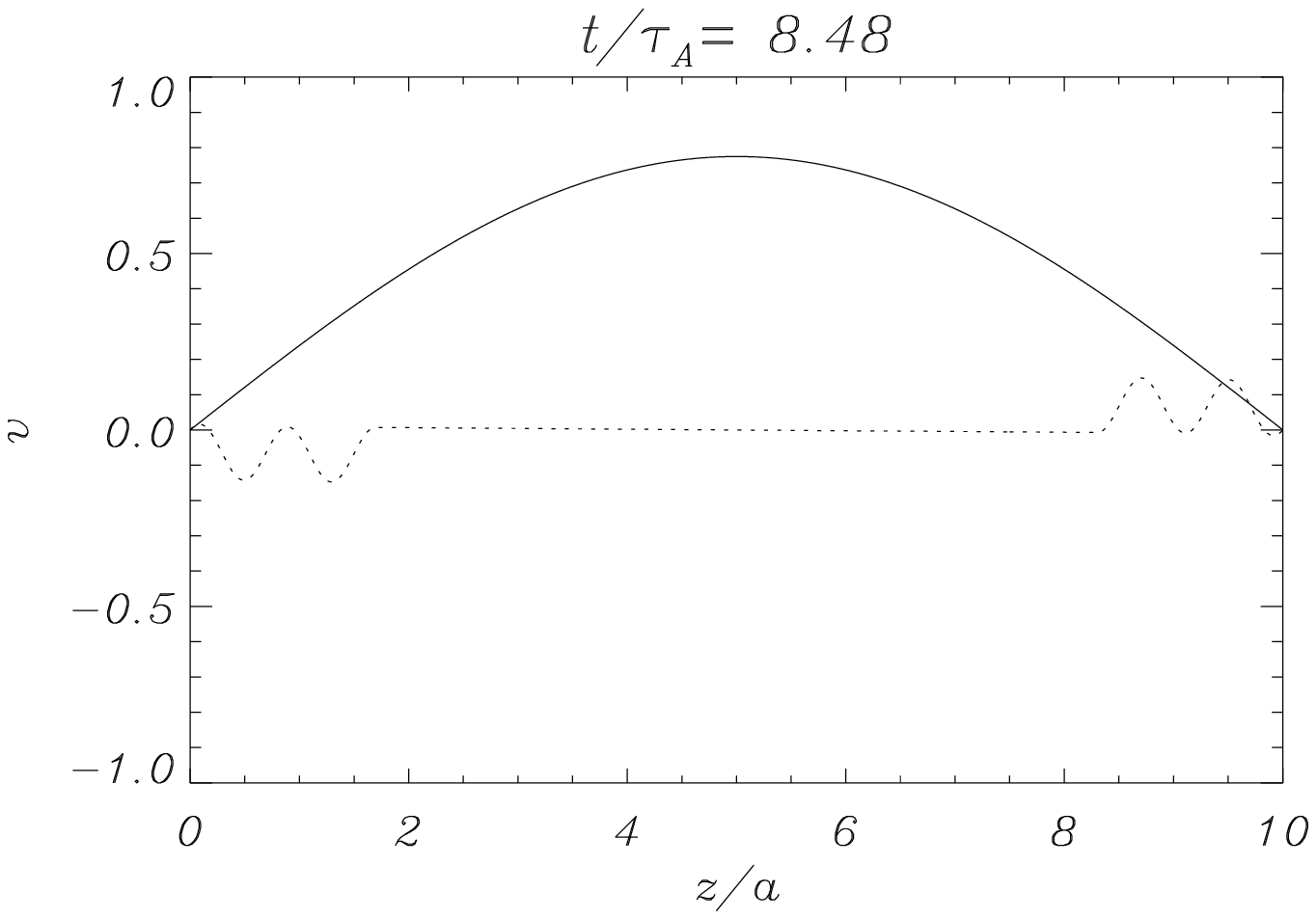}}
\center{\includegraphics[width=7cm]{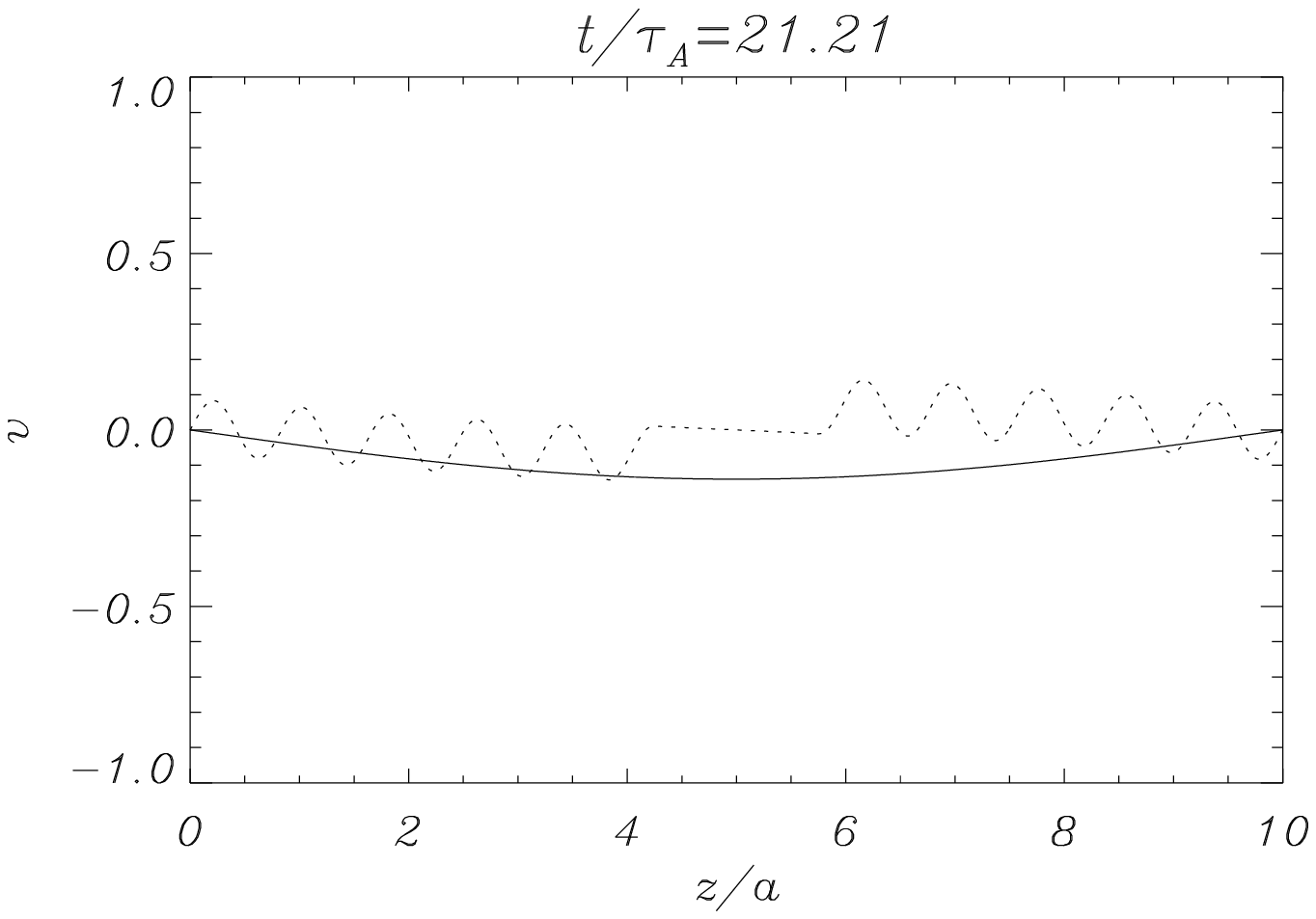}}
\center{\includegraphics[width=7cm]{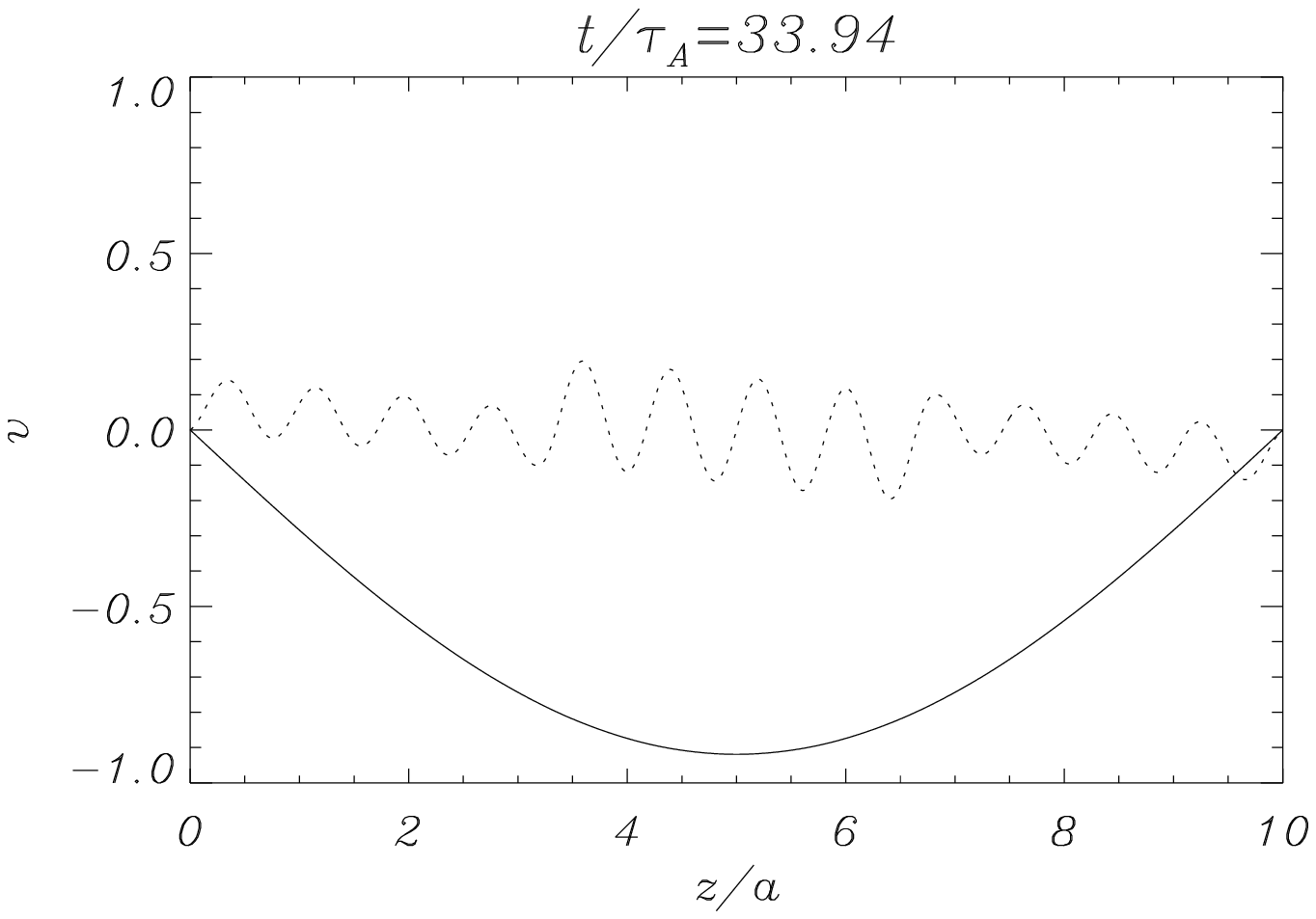}}
\caption{\small Time evolution of the standing fast excitation given by
Eqs.~(\ref{v_xstanding})-(\ref{v_zstanding}). The continuous line represents the $v_x$ component while
the $v_z$ component is plotted with a dotted line (its amplitude has been
multiplied by a factor 10 for visualisation purposes). For this simulation $k
a=\pi/10$, $v_0/v_{\rm A}=1$, $k_x a=1.5$ and $c_{\rm s}/v_{\rm A}=0.2$.}
\label{standing0plot} \end{figure}

Line-tying conditions are applied at the two loop footpoints. We choose the
longitudinal wavenumber that excites the fundamental standing fast MHD mode,
i.e., $k=\pi/L$. The results, represented in Figure~\ref{standing0plot} manifest
the generation of slow modes at the footpoints (top panel). The fast mode drives
slow modes which move in the direction of the loop apex, located at $z=L/2$ (see
middle panel). At a certain time (when $t\simeq L/2c_{\rm s}$) the interference
between the opposite propagating slow waves is produced. Since the two slow
waves are identical but travelling in opposite directions a quasi-standing
pattern is visible (see the profile of $v_z$ around $z=5a$ in the bottom panel).
Eventually, slow modes reach the opposite footpoint and get reflected. Under
such conditions the inverted process takes place, i.e., the slow mode generates
a fast mode. Nevertheless, this problem would take us too far and it is out of
the scope of this work. However, it is worth to note that depending on the
equilibrium and wave parameters the generated slow modes might match the
frequency of a standing slow eigenmode of the loop. This means that since the
slow mode is driven by the fast standing mode, the frequencies of the slow
standing and fast standing waves will be basically the same and the modes will
have a highly mixed nature. This takes place around the
``avoided crossings" in the dispersion diagram. 

\begin{figure}[!ht]
\center{\includegraphics[width=7cm]{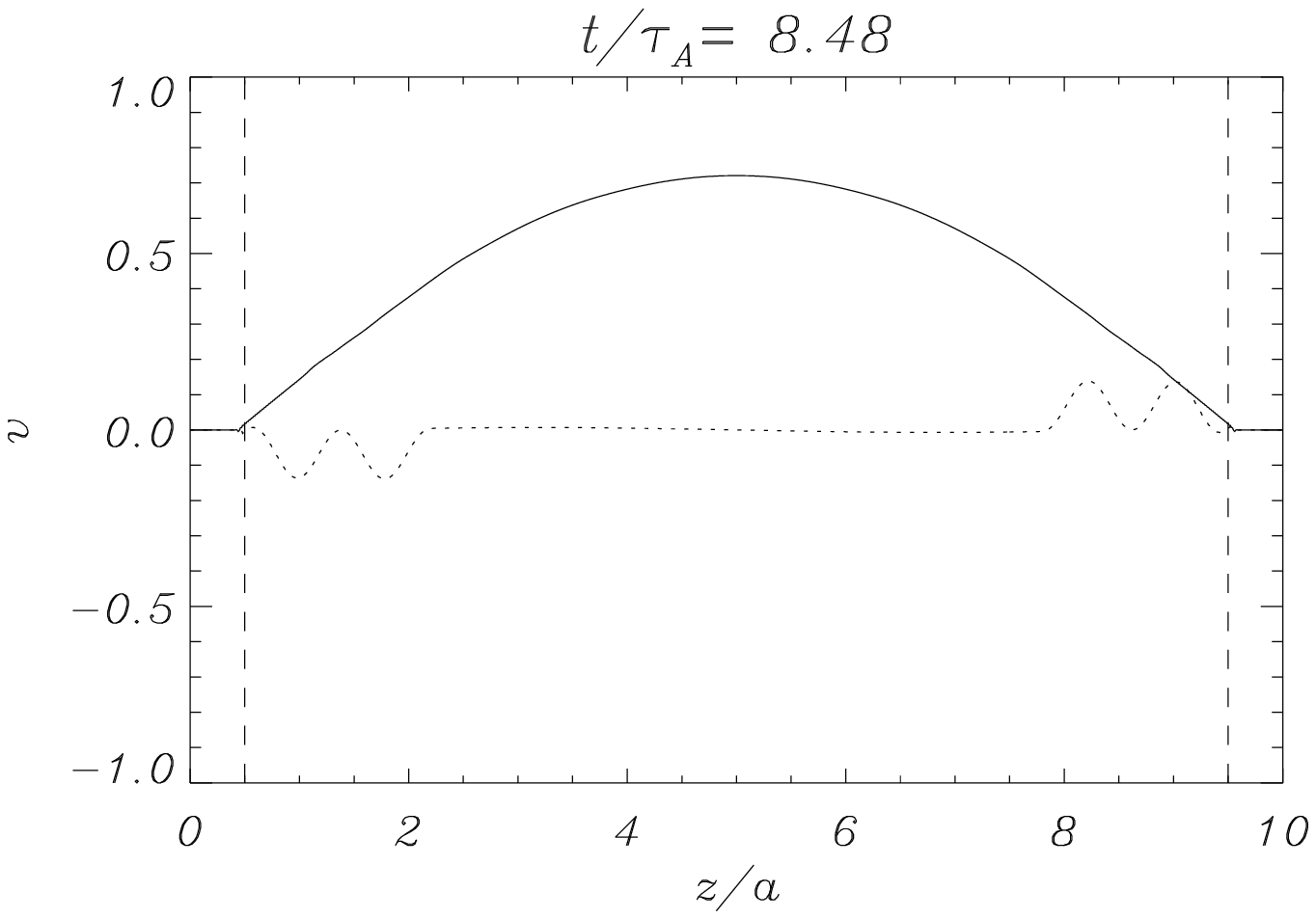}}
\center{\includegraphics[width=7cm]{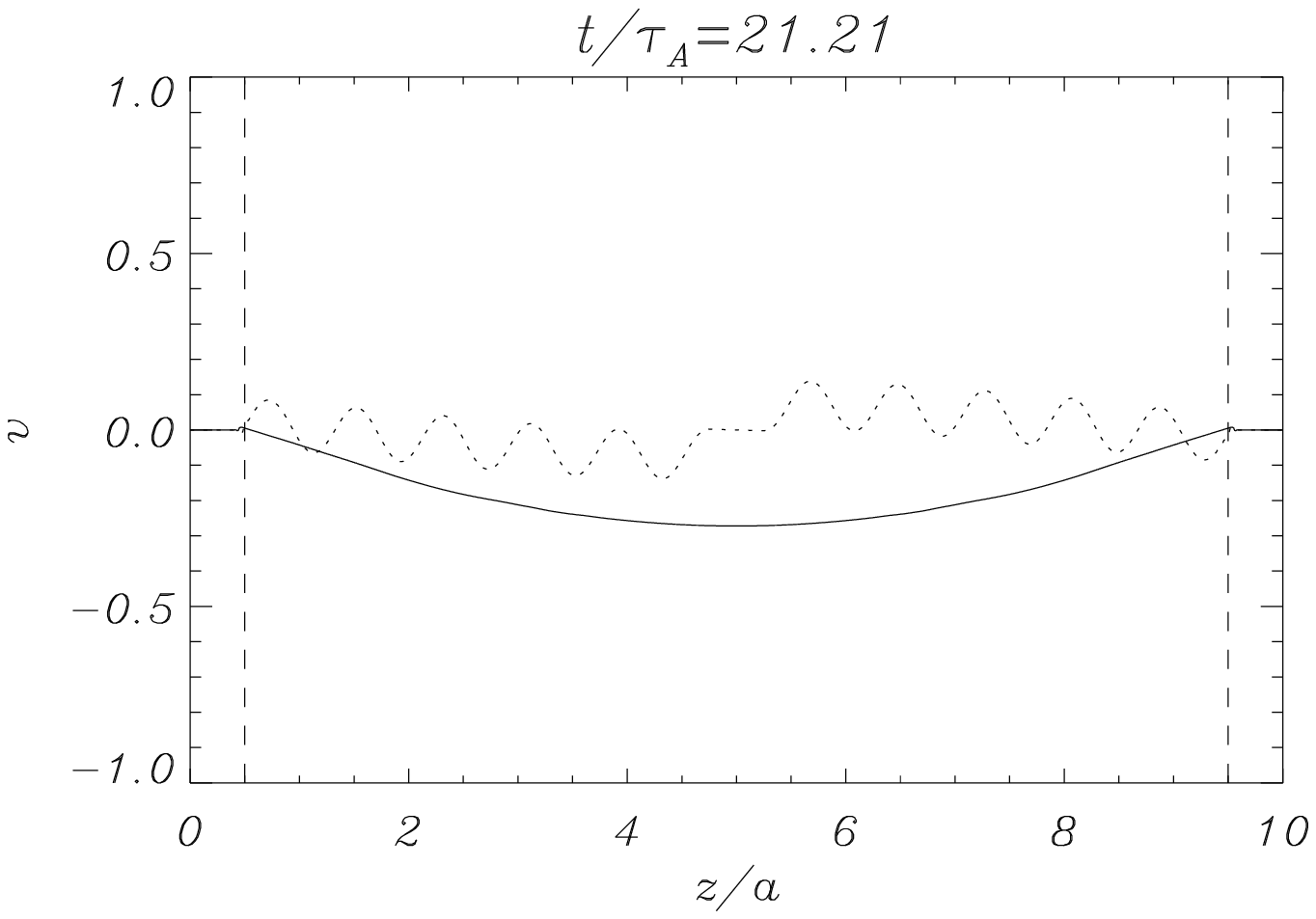}}
\center{\includegraphics[width=7cm]{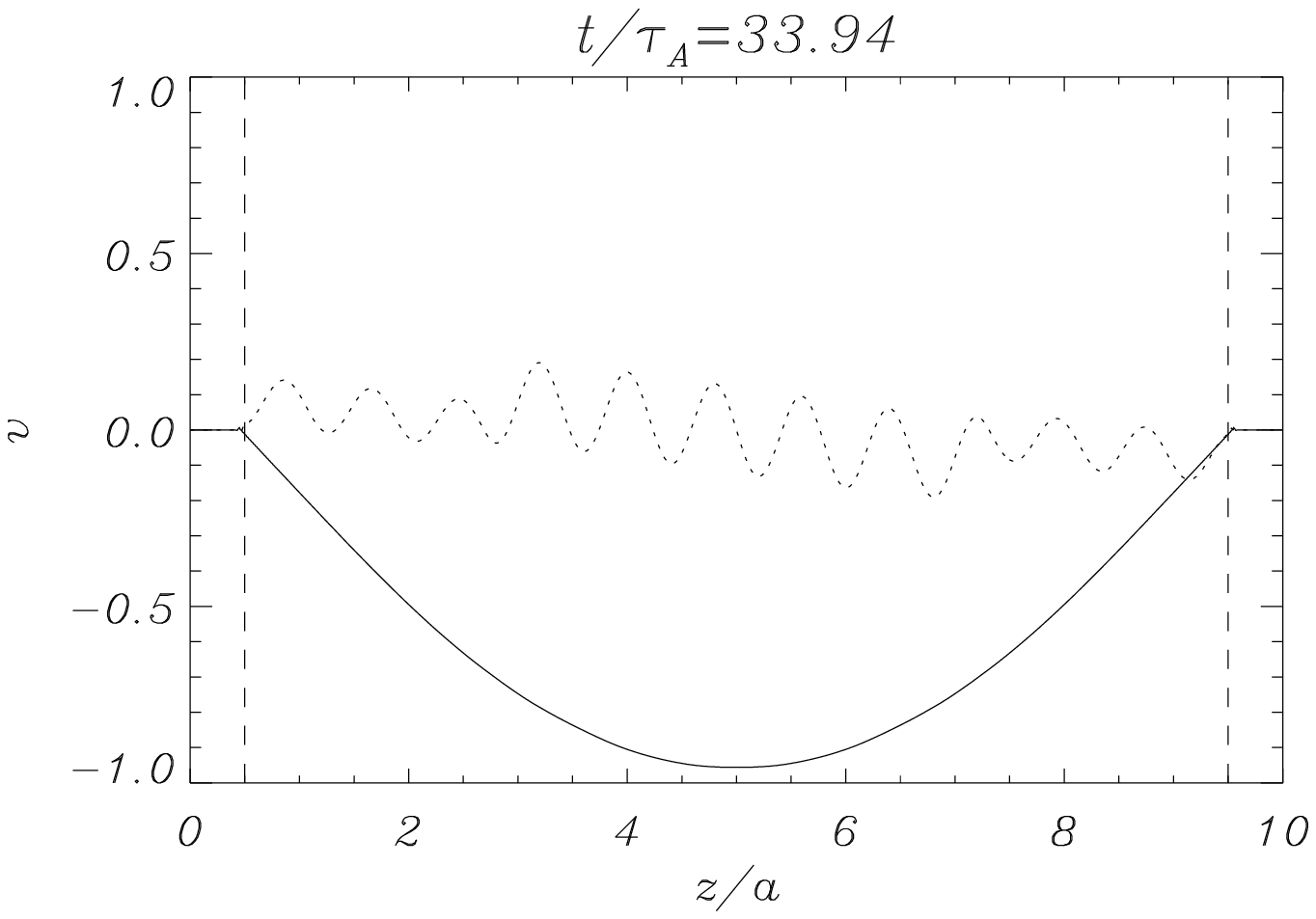}}
\caption{\small Time evolution of the standing fast excitation given by
Eqs.~(\ref{v_xstanding})-(\ref{v_zstanding}) with the density model given by
Eq.~(\ref{densphot}). The continuous line represents the $v_x$ component while
the $v_z$ component is plotted with a dotted line (its amplitude has been
multiplied by a factor 10 for visualisation purposes). The vertical dashed lines
represent the locations of the transition between the corona and photosphere. For this simulation $k
a=\pi/9$, $v_0/v_{\rm A}=1$, $k_x a=1.5$, $c_{\rm s}/v_{\rm A}=0.2$, $\rho_{\rm
ph}/\rho_{\rm c}=10^8$,
$w/a=0.25$.}
\label{standing1plot} \end{figure}

\subsubsection{Sharp photosphere-corona transition }

From the previous results we might think that the generation of slow modes takes
place only when rigid boundary conditions are imposed. In order to address this
question, instead
of line-tying conditions, a density transition between the corona and the
photosphere is considered. As density profile we use the following idealised
model,
\begin{eqnarray} \label{densphot} \rho=\left(\rho_{\rm ph}-\rho_{\rm c}\right)
e^{-\left(\frac{z}{w}\right)^4}+\left(\rho_{\rm ph}-\rho_{\rm c}\right)
e^{-\left(\frac{z-L}{w}\right)^4}+\rho_{\rm c}.
\end{eqnarray}

\noindent The parameter $w$ measures the width of the transition between the
photosphere, with density $\rho_{\rm ph}$, and corona, with density $\rho_{\rm
c}$. The density contrast is set to $\rho_{\rm  ph}/\rho_{\rm c}=10^8$.
Non-reflecting conditions are imposed at the boundaries, now placed below the
photosphere. The details of the model used to represent the transition
photosphere-corona are not important since our focus is on the generated slow
modes. The equilibrium gas pressure is set to the same constant value in the
corona and  photosphere to have magnetohydrostatic equilibrium.

The results of the time-dependent simulation are plotted in
Figure~\ref{standing1plot}. There are some differences with respect to the
behaviour found for the perfectly reflecting boundaries. For example, the system
now allows the energy to escape through the photosphere, i.e., there is a
transmitted fast (and slow wave). This means that the energy can leak through the
photosphere but this is not visible in Figure~\ref{standing1plot} because the
amplitude is very small. Nevertheless, the primary result here is that a similar
slow mode generation in comparison with the case of reflecting boundaries is
produced in the coronal part (compare Figures~\ref{standing0plot}
and~\ref{standing1plot}). The results are qualitatively similar, thus, the
overall conclusion is that nonuniformity causes the coupling between fast and
slow modes in a similar way as line-tying conditions.

\section{Link with observations}

It is interesting to look for signatures of the mode coupling studied in this
work in the observations. Recently, \citet{verwetal10}, have reported, using
combined observations from TRACE and EIT/SoHO, intensity fluctuations at one loop
footpoint with basically the same period, around 40 min, as the fast transverse
oscillation of the loop. This long periodicity indicates that the origin of these
density oscillations is most probably not photospheric. \citet{verwetal10} suggest
that the intensity oscillations are due to variations in the line of sight column
depth produced by the changes in the loop inclination as it oscillates with the
transverse kink mode. This would explain the coincidence in periods, and also in
the damping times.

\begin{figure}[!ht]
\center{\includegraphics[width=9cm]{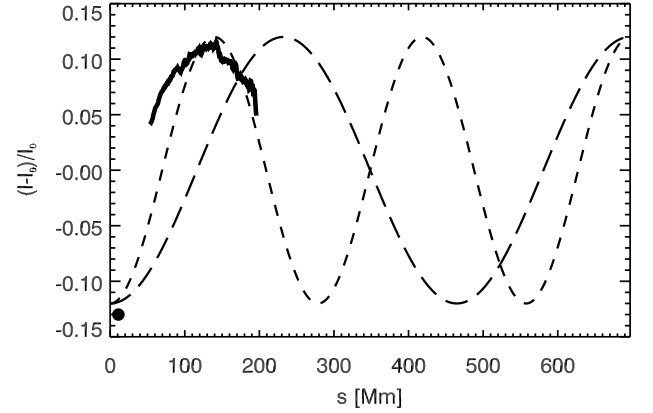}} \caption{\small
Relative profile of intensity variations seen in a
transversely oscillating loop studied by \cite{verwetal10} using TRACE as a
function of distance along the loop. The solid circle is the measurement of the
intensity variation at the loop foot point. The long-dashed and dashed curves
are functions -0.13$\sin(3\pi/L)$ and -0.13$\sin(5\pi/L)$, respectively,
representing a possible third or fifth standing harmonic of the slow mode.
}
\label{verw} \end{figure}

According to our study an alternative explanation for the intensity variations
reported by \citet{verwetal10} is done in terms of the coupling between fast and
slow waves due to the line-tying conditions. The similarity of the periods of
the transverse mode and the intensity fluctuations is explained by our model
where the excited slow wave, and the corresponding intensity oscillation, has
the same periodicity as the transverse loop motion.  \citet{verwetal10} show
intensity variations associated with a transverse oscillation in a large loop of
690$\pm$60 Mm length. The oscillation, which is a horizontally polarised
fundamental kink mode, has a period of $P$=2418$\pm$5 s. Therefore, the phase
speed is equal to $v_\mathrm{ph}$=580$\pm$50 km\,s$^{-1}$. For a thin loop in
the long wavelength limit, this phase speed tends to the kink speed,
$c_\mathrm{k}$, of a cylindrical tube. If the intensity oscillation is associated
with a slow magnetoacoustic mode with the same periodicity as the kink mode,
then its wavelength is equal to 
$\lambda_\mathrm{s}$=$c_\mathrm{s}P$=$(c_\mathrm{s}P/v_\mathrm{ph})2L$. For a
temperature between 0.9 and 1.8 MK, the coronal sound speed is in the range
144-198 km\,s$^{-1}$. Hence, $\lambda_\mathrm{s}$=(0.30$\pm$0.07) $2L$. For a
standing mode, this translates into a wavenumber
$n$=$v_\mathrm{ph}/c_\mathrm{s}$=$2L/\lambda_\mathrm{s}$=3.3$\pm$1.0. The EIT
observations suggests that the intensity variations at the two foot points of
the loop oscillate in anti-phase. Hence, the mode, if standing, should be an odd
harmonic. Figure~\ref{verw} shows the relative profile of intensity variations
as a function of loop distance. The relative intensity has an amplitude of
approximately 13\%, which translates into a density perturbation with an
amplitude, $|\rho_1/\rho|_\mathrm{max}$=6.5\%. However, it is clear that a third
or fifth harmonic could fit the observed intensity variation if we take into
account that most likely the slow mode is still evolving towards a standing wave
pattern. We have measured the full intensity profile as a function of distance
along the loop using EIT. Nevertheless, this has large uncertainties attached to
it because of the lack of resolution and line-of-sight confusion. Also, the loop
may be longitudinally structured in temperature. The EIT measurement hints at
the presence of a fifth harmonic. \citet{verwetal10} show that the displacement
amplitude, $\xi_0$, can be modelled by a ten degree inclination of the loop. As
the loop has a height of 236 Mm, this means that $\xi_0$=41 Mm. Hence, the
velocity amplitude $V_0$=$\xi_0\omega$=106 km\,s$^{-1}$. Using
Eq.~(\ref{denstslowstand}), and the estimated value for the Alfv\'en speed,
$v_\mathrm{A}$=410$\pm$40 km\,s$^{-1}$, we find
$|\rho_1/\rho|_\mathrm{max}$=5$\pm$3\%. This is consistent with the observed
density amplitude (around 6.5\%).


In another work, \cite{verwetal09} also reported intensity oscillations
associated with a kink mode seen by EUV on board of STEREO and interpreted these
as resulting from variations in the line-of-sight column depth due to the loop
showing a varying aspect to the observer. The measured relative density
amplitude is 2.5\%, and located around to loop top. Can this observation be
explained by a slow mode instead? In that observation, the key parameters that
were measured are $L$=340$\pm$15 Mm, $v_\mathrm{ph}$=1100$\pm$100 km\,s$^{-1}$,
$V_0$=35$\pm$9 km\,s$^{-1}$ and $v_\mathrm{A}$=800$\pm$100 km\,s$^{-1}$. For the
peak temperature of the EUV 171\AA \, bandpass of 0.9 MK, $c_\mathrm{s}$=144
km\,s$^{-1}$.  The slow mode would have a wavenumber $n$=7$\pm$1 (in the case of
a standing mode), corresponding to a wavelength of about
$\lambda_\mathrm{s}$=90$\pm$10 Mm. Again, using Eq.~(\ref{denstslowstand}), we
calculate $|\rho_1/\rho|_\mathrm{max}$=0.4\%, which is negligibly small.
Therefore, for this observation, the slow mode is not able to explain the
observed intensity variations. The main difference with respect to the case
studied by \citet{verwetal10} is that the density perturbations are located
around the loop top instead of the loop footpoints.

\section{Discussion and conclusions}

In this work we have investigated the possible effects of linear coupling
between fast and slow modes that takes place due to the reflection at the
photosphere. We have shown, by solving the reflection problem, that the
transverse motion of the loop produces slow MHD waves as long as the fast wave
is obliquely incident on the boundary. These slow waves are manifested as
propagating density fluctuations with the same frequency as the standing
transverse mode and are generated at the loop footpoints. The wavelength of
these slow modes is basically given by $k_s\simeq\omega/c_{\rm s}$, and since
$v_{\rm A}>c_{\rm s}$ their wavelength is smaller than that of the corresponding
fast standing transverse mode. Moreover, we have derived using the reflection
coefficient of the slow mode, simple analytical expressions that relate the
transverse displacement at the loop top with the amplitude of the density
fluctuations produced at the footpoints. The coupling between fast and slow is
proportional to the plasma-$\beta$, indicating that under coronal conditions it
will be weak in general. However, it is proportional to the amplitude of
oscillation of the fast mode, which can be large in some cases, and close to the
footpoints, i.e., close to the photosphere, the ratio between the sound and
Alfv\'en speeds can increase. Ideally, from the properties of the reflected slow
modes, we could have indirect information about the real line-tying conditions
in coronal loops and realistic values of the sound speed near the footpoints. 

The generated slow waves can eventually form a standing pattern along the loop.
Thus, mode coupling might provide an excitation mechanism of standing slow
modes, different from the driven photospheric origin usually emphasised in the
literature. However, since the sound speed is assumed to be smaller than the
Alfv\'en speed, the time required for the slow mode to travel along the loop and
reflect at the footpoint is much larger than that of the fast MHD mode. This
means that slow standing modes will need much more time than fast standing modes
to build up. Nevertheless, if the wavelength of the slow modes is very short
they might be damped by thermal conduction before the standing wave is formed. 

It has been shown that in the observations analysed by \citet{verwetal10} there
are possible fingerprints of the coupling between fast and slow modes due to
line-tying, providing an alternative explanation to the effects of integration
along the line of sight proposed by these authors. However, in the observations
studied by \citet{verwetal09} the mechanism is unable to explain the intensity
oscillations. One of the reasons might be that the density changes observed in
this event are reported around the loop apex, while the estimations are based on
propagating waves around the footpoints. In the case of the standing pattern the
density fluctuations can be much larger. Further analysis of other observations
will help to test the linear coupling as an operative mechanism in coronal
loops. 

It is worth to mention that the mode coupling discussed in the work is a purely
linear effect. It is unrelated to the nonlinear coupling between fast and slow
modes due to the ponderomotive force
\citep[see][]{hollweg71,rankinetal94,terrofm04}, or the parametric coupling
studied by \citet{temurietal02,temurietal05}. It is also different from mode
conversion that takes place when  $c_{\rm s}=v_{\rm A}$. Line-tying boundary
conditions are the responsible ingredients of the coupling, but we have shown
that they are not the only way in which fast and slow waves may couple. A sharp
transition between the corona and the photosphere produces the generation of slow
modes, i.e., a change in the properties of the medium, for example the transition
region, leads inevitably to the coupling. 

Our theoretical model is based on a Cartesian geometry without a  density
enhancement representing a loop. For this reason, it is necessary to improve the
model by including a slab or a cylinder that mimics the effect of a dense
magnetic tube. This will complicate the mathematical problem of the reflection
at the boundary, and it might be difficult to derive simple analytical
expressions. For example, in a cylindrical model, the eigenfunctions of the slow
and fast MHD waves have different radial dependencies and might share the same
azimuthal wavenumber (playing basically the role of $k_x$ in the present work).
This issue will be addressed in a future study.







\begin{acknowledgements} J.T. acknowledges the Universitat de les Illes Balears for a
postdoctoral position and the financial support received from the Spanish MICINN and
FEDER funds (AYA2006-07637). J.T. thanks Ram\'on Oliver and Roberto Soler for
useful comments and suggestions. J.A. is supported by an International Outgoing Marie Curie
Fellowship within the 7th European Community Framework Programme. J.A. also acknowledges
support by the Fund for Scientific Research - Flanders. E.V. acknowledges financial
support from the UK Engineering and Physical Sciences Research Council (EPSRC) Science
and Innovation award.\end{acknowledgements}


\appendix
\section{}\label{app}

We use the linearised MHD equations of the momentum, induction, energy and 
continuity equation. Assuming Fourier analysis in the $x-$ direction we have:
\begin{eqnarray}
\frac{\partial v_x}{\partial
t}&=&\frac{1}{\rho}\left(-i k_x p_1-i k_x \frac{B_0}{\mu} b_z+ \frac{B_0}{\mu} \frac{\partial b_x}{\partial
z}\right),\\  
\frac{\partial v_z}{\partial
t}&=&-\frac{1}{\rho}\frac{\partial p_1}{\partial
z},\\ 
\frac{\partial b_x}{\partial
t}&=&B_0\frac{\partial v_x}{\partial
z},\\ 
\frac{\partial b_z}{\partial
t}&=&-B_0 i k_x v_x,\\ 
\frac{\partial p_1}{\partial
t}&=&-\gamma p \left(i k_x v_x+ \frac{\partial v_z}{\partial
z}\right),\\
\frac{\partial \rho_1}{\partial
t}&=&-\rho \left(i k_x v_x+ \frac{\partial v_z}{\partial
z}\right).\label{appx}  \end{eqnarray}
In these equations $B_0$ is the equilibrium magnetic field and $\rho$ and $p$ the
equilibrium density and gas pressure, respectively. The rest of the variables
correspond to the perturbed magnitudes. To eliminate the imaginary complex
numbers we simply define \begin{eqnarray} v^{*}_x=i\, v_x,\\ b^{*}_x=i\, b_x.
\end{eqnarray} With this transformation the time-dependent equations are solved
using standard numerical techniques.  

The dispersion relation, given by Eq.~(\ref{omdisper}), is easily derived
assuming in the previous equations a temporal dependence of the form $e^{i\omega
t}$ and a spatial dependence with the $z-$coordinate of the form $e^{i k_z z}$.
We have used the following standard definitions for the Alfv\'en and sound
speeds, \begin{eqnarray} v_\mathrm{A}&=&\frac{B_0}{\sqrt{\mu \rho}},\\
c_\mathrm{s}&=&\sqrt{\gamma \frac{p}{\rho}}. \end{eqnarray}

\end{document}